\documentclass[a4paper,
               keeplastbox,   
               ]{jacow}
\usepackage{pdfpages,multirow,ragged2e} %
\usepackage{pgfplots}
\pgfplotsset{compat=1.14}
\makeatletter%
	\ifboolexpr{bool{xetex}}
	 {\renewcommand{\Gin@extensions}{.pdf,%
	                    .png,.jpg,.bmp,.pict,.tif,.psd,.mac,.sga,.tga,.gif,%
	                    .eps,.ps,%
	                    }}{}
\makeatother

\ifboolexpr{bool{xetex} or bool{luatex}} 
 {}                                      
 {\usepackage[utf8]{inputenc}}           

\usepackage[USenglish]{babel}

\ifboolexpr{bool{jacowbiblatex}}%
 {%
  \addbibresource{jacow-test.bib}
  \addbibresource{biblatex-examples.bib}
 }{}
\begin{document}

\title{Sirius Digital Low-Level RF Status}

\author{A. P. B. Lima\thanks{andre.lima@lnls.br}, F. K. G. Hoshino, C. F. Carneiro, R. H. A. Farias\\Brazilian Synchrotron Laboratoy (LNLS), Campinas, Brazil\\
		A. Salom, CELLS – ALBA, Barcelona, Spain}
	
\maketitle

\begin{abstract}

Sirius is a Synchrotron Light Source Facility based on a 4th generation low emittance storage ring. The facility is presently in the final assembly phase in Campinas, Brazil, and comprises a \SI{3}{GeV} electron storage ring, a full energy  Booster and a \SI{150}{MeV} Linac. The Booster RF system is based on a single 5-cell RF cavity driven by a \SI{45}{kW} amplifier at \SI{500}{MHz} and is designed to operate at \SI{2}{Hz} repetition rate. The storage ring will initially use one normal conducting 7-cell RF cavity and, in the final configuration, will comprise two superconducting cavities, each one driven by a \SI{240}{kW} RF amplifier. A digital LLRF system based on ALBA DLLRF has been designed and commissioned to control both normal conducting and future superconducting RF configurations. The first DLLRF system is already in operation in the injector Booster. This paper presents an overview of the performance of the control loops in high power operation and the results of the Booster DLLRF commissioning. It also shows the integration of DLLRF with the RF system as well as other utilities of the system such as automatic start-up, conditioning, fast interlocks and post-mortem analysis.
\end{abstract}

\section{INTRODUCTION}

Sirius is the new 4th generation \SI{3}{GeV} low emittance light source under construction at the Brazilian Synchrotron Light Laboratory (LNLS). The storage ring natural emittance of \SI{0.25}{nm.rad} is reached with twenty 5BA lattice cells and it can be further reduced to \SI{0.15}{nm.rad} as insertion devices are added. The Sirius injector consists of a \SI{150}{MeV} linear accelerator and a full energy synchrotron Booster installed in the same tunnel and concentric with the storage ring. The installation and commissioning of Linac was finished in 2018. Currently, the Booster is being commissioned in high energy~\cite{Rodrigues:IPAC2019-TUPGW003}.

In its final configuration the storage ring RF system will include two superconducting cavities, each one driven by \SI{240}{kW} solid state power amplifiers (SSPA) at \SI{500}{MHz}. However, a 7-cell RF cavity will be used for commissioning and initial beamlines operation, driven by a \SI{120}{kW} SSPA. The Booster RF system is based on a 5-cell RF cavity driven by a \SI{45}{kW} SSPA.

A digital low level RF (DLLRF) capable of controlling these three types of cavities has been designed and assembled based on the ALBA's mature design to achieve \SI{0.1}{\%} amplitude and $\SI{0.1}{^\circ}$ phase stability under normal operating conditions.

ALBA's DLLRF is based on commercial-off-the-shelf hardware and different versions of that design are currently in operation in other facilities like MAX-IV, SOLARIS and Diamond (DLS)~\cite{Salom:IPAC2017-THPAB135, borowiec2018operational, Gu:IPAC2017-THPAB152}. Sirius DLLRF uses a commercial device, a PicoDigitizer from Nutaq~\cite{Salom:IPAC2019-THPTS060}, as the central core of the system. The PicoDigitizer includes a Virtex-6 FPGA motherboard with ADC and DAC FMC boards. Communication with the FPGA motherboard is done via Gigabit-Ethernet and accessing an embedded Linux processor of the FPGA. 

\section{SIRIUS DLLRF HARDWARE}

\begin{figure}[!htb]
   \centering
   \includegraphics*[width=1\columnwidth]{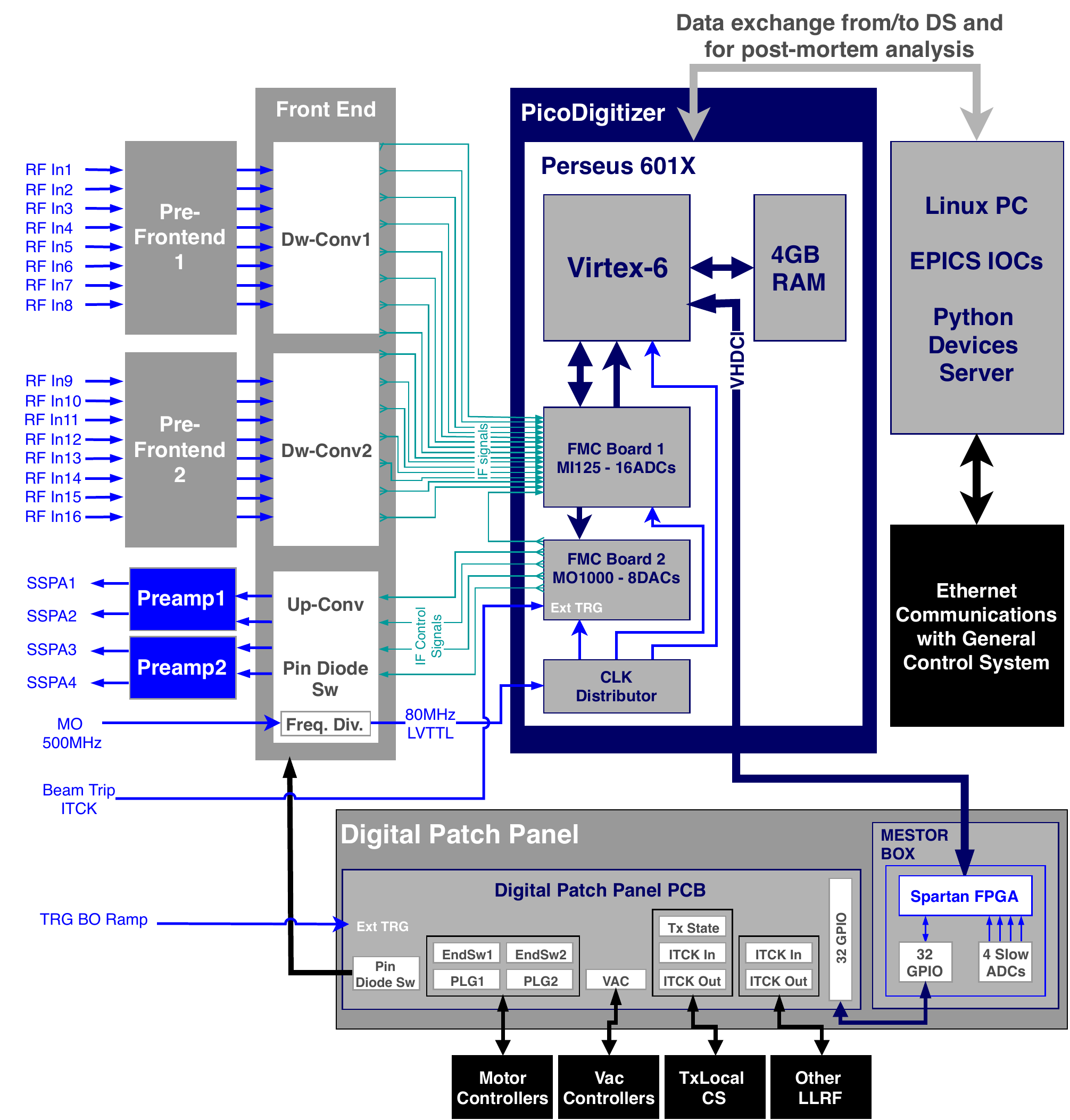}
   \caption{Main hardware components of Sirius DLLRF.}
   \label{fig:main_hardware_components}
\end{figure}

Fig.~\ref{fig:main_hardware_components} shows the main hardware components of the DLLRF and the interaction between different RF and digital signals of the subsystems, such as Front Ends, Digital Patch Panel, PicoDigitizer and Preamplifier. Sirius DLLRF hardware includes:
\begin{Itemize}
  \item The PicoDigitizer that houses the FPGA motherboard and ADC/DAC FMC boards.
  \item Mestor interface with digital GPIO bus.
  \item Front ends for up and down conversion of RF signals.
  \item A clock generator based on a frequency divider ensuring synchronization with Master Oscillator (MO).
  \item Patch Panels with connector and level adapters between DLLRF and RF plant subsystems.
\end{Itemize}

\begin{figure}[!tbh]
    \centering
    \includegraphics*[width=1\columnwidth]{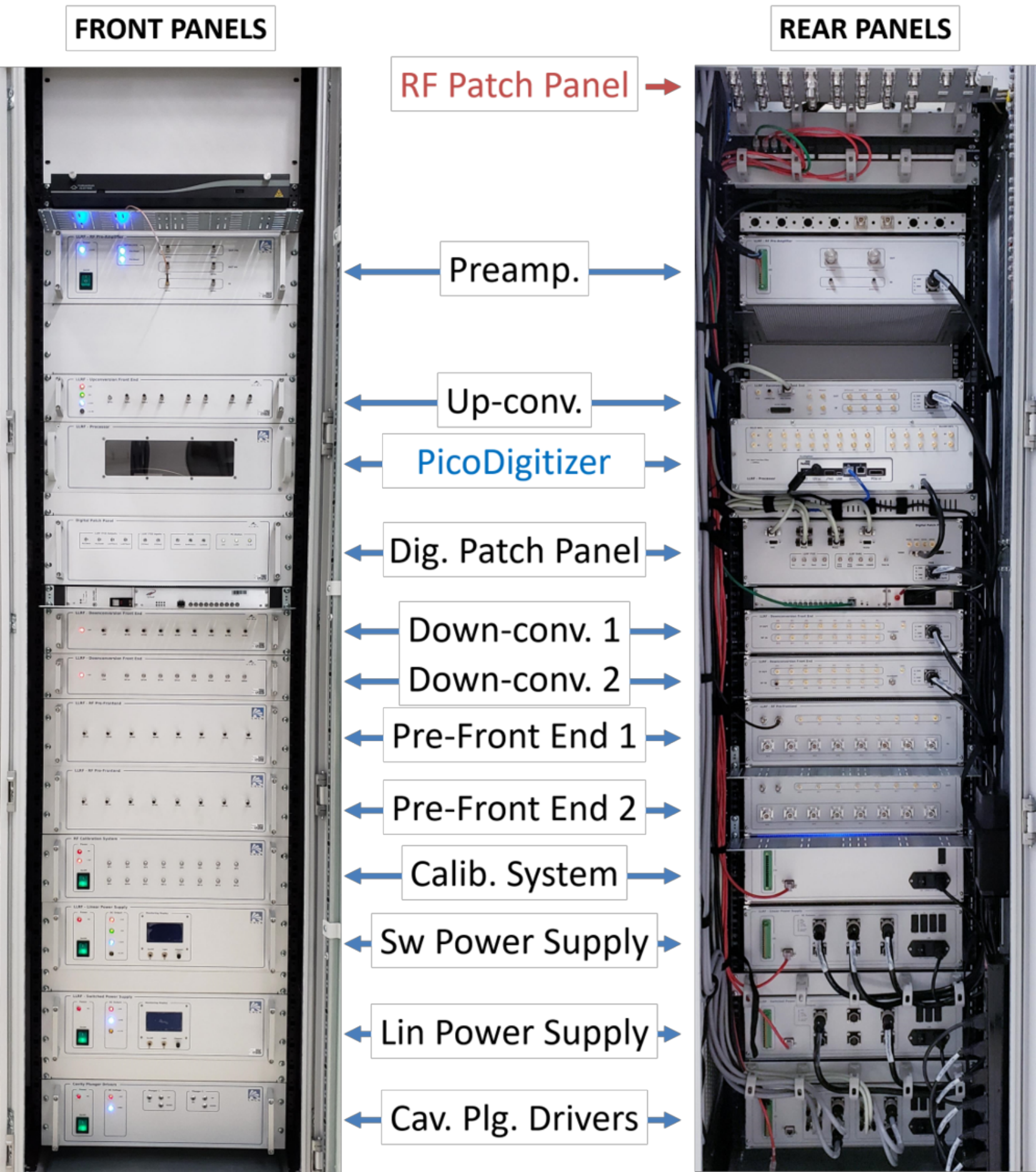}
    \caption{Front and Rear Panels of the Sirius Booster DLLRF components.}
    \label{fig:Booster_rack}
\end{figure}

Fig.~\ref{fig:Booster_rack} shows the front and rear panels of the main hardware components of the Sirius DLLRF.

The PicoDigitizer has a Virtex-6 SX315T FPGA and two FMC boards: a 16-channel 14-bit \SI{125}{MSPS} ADC board and a 8-channel 16-bit \SI{250}{MSPS} DAC board. The resolution of the ADCs are better than \SI{0.06}{\percent} rms in amplitude and $\SI{0.04}{^\circ}$ rms in phase and the signal to noise ratio is better than \SI{70}{dB}~\cite{Gu:IPAC2017-THPAB152}. It also includes an interface module, the Mestor expansion board, with a 32 bit GPIO bus that is used for digital interfaces with other subsystems like motor controller, vacuum controllers, timing and control system. The Digital Patch Panel adapts the level of the GPIO bus signals (LVTTL) to whatever digital level required by the other subsystems: dry contact, open collector, TTL, PLC \SI{24}{V}, etc.

\subsection{Front Ends and Timing System}

The Down-Conversion Front End transforms the \SI{500}{MHz} signals into \SI{20.83}{MHz} Intermediate Frequency (IF) signals by mixing with \SI{479.16}{MHz} Local Oscillator (LO) signal. The IF signals are sampled by the PicoDigitizer ADCs at four times the IF frequency, condition required to perform an IQ-demodulation.

The IF control signals provided by the PicoDigitizer DACs is mixed with LO, filtered and amplified to generate the \SI{500}{MHz} RF signals in the Up-Conversion Front End. Also, there is a RF PIN switch in the Up-Conversion unit that can be triggered by an interlock. These switches are opened in less than \SI{1}{\micro s} if an interlock is detected. 

The IF frequency required to generate the LO of \SI{479.16}{MHz} is provided by a Frequency Divider (Valon Frequency Divider 3010). The input of this device is the MO and it provides up to three outputs that can be equal to the input frequency divided by integers between 1 and 32. The Valon Frequency Divider is used to provide the clock signal for the FPGA and ADCs/DACs equals to the MO divided by 6.

The Front End supplies SMA test points in the front panels as shown in Fig.~\ref{fig:Booster_rack} for testing purposes. They are connected through directional couplers to the main RF signals of the DLLRF.

\subsection{Pre-Front End}

The Pre-Front End is the first stage of the RF signal path. There are two Pre-Front End units in each DLLRF system to handle all 16 RF channels available. Each unit consists of 8 isolated input channels and its main function is to adjust the \SI{500}{MHz} RF signals to ensure that the RF power does not exceed the maximum input level of the Front End modules. The front panel provides a coupled signal for each input channel for monitoring purposes. In addition, a DC voltage signal is provided for the Sirius Personal Protection System which can be used to inform when there is RF power inside the tunnel. 

\subsection{Calibration System}

The Calibration System hardware is designed to measure the RF power of an incoming signal and provide a measurement of the power level in \SI{}{dBm}. It has 16 RF input ports and its main function is to provide EPICS variables with the measurement of the power level at each of these ports.

 \textit{Mini-Circuits}'s \textit{ZX47-40LN} RF power detector is used to measure the power of each RF signal. A 12-bit \textit{Analog Devices} ADC \textit{AD7923} and a BeagleBone Black (BBB) single-board computer are used to provide the digitized readings from the detector in EPICS variables.

The LLRF Calibration System was designed to assist in the calibration procedure of the RF signals. The main task is to convert the PicoDigitizer ADCs readings in \SI{}{mV} to RF power units, such as \SI{}{dBm} or \SI{}{W}. The Calibration System is ready to measure the RF signals from the Pre-Front End monitoring channels. A Python script is responsible for adjusting the RF system variables and reading back the RF power from the PicoDigitizer in \SI{}{mV} and from the Calibration System in \SI{}{dBm} for a range of power level steps. A fourth-degree polynomial curve is used to fit these two data to find the best conversion function.

\subsection{Preamplifier}

The preamplifier unit is the first amplification stage of the RF Drive signals from the up-conversion Front End before being delivered to the SSPA. This device have two isolated channels with \SI{20}{dB} gain and can provide up to \SI{2}{W} of output power in each channel. In addition, it includes a RF switch that can be triggered by the RF Interlock System.

\subsection{RF Patch Panel and RF cables}

The RF patch panel is meant to facilitate the cable connections of Input and Output RF signals between the DLLRF Rack and other RF subsystems, such as SSPA, RF Cavity and Circulator. Double shielded RG316 coaxial cables are used to connect RF crates units inside the DLLRF Rack to ensure a better signal integrity and isolation.

\section{SIRIUS DLLRF FIRMWARE}

Sirius DLLRF firmware has many features that has been useful for machine operation~\cite{Salom:IPAC2019-THPTS060}, which includes:
\begin{Itemize}
  \item Cavity configuration according to its type such as Superconducting and Multi-cell or Single-cell Normal Conducting cavities.
  \item IQ Digital Demodulation.
  \item PI Loops for Cavity Voltage Control in IQ (rectangular) or Polar loops in case amplitude and phase loops needs to be adjusted independently.
  \item Phase Shifters and Gain control on each DAC's outputs and ADC's inputs.
  \item Booster Ramping at \SI{2}{Hz} cycle.
  \item Tuning and Field Flatness Loop for cavity resonance and field distribution control.
  \item Automatic Start-up for a faster recovery of the RF System.
  \item Automatic Conditioning to speed up the conditioning process of the cavity according to the vacuum level.
  \item Fast Interlocks and Fast Data Logger for Machine Protection and post-mortem analysis.
\end{Itemize}

\section{DLLRF Control System}

The integration of the LLRF system to the machine's control platform requires an EPICS IOC. The development of this IOC started from a basecode provided by DLS and is still underway. Although several customization have been made in recent months and there is already implemented an operational version of the IOC, many improvements are still pending.

Concerning the GUI screens for the DLLRF operation, current implementation evolved from EDM screens also created at DLS. Due to the convenience of standardizing the GUI of the operation programs a complete migration to CS-Studio (BOY) was undertaken after a few months of development. As the commissioning phase advances, LLRF screens will be modified to become more intuitive in such a way to address all operation requirements. 

\section{DLLRF PERFORMANCE EVALUATION}

Currently, the commissioning of the Booster 5-cell RF cavity is in its final steps and \SI{45}{kW} CW forward power has been reached.

\begin{figure}[!tbh]
    \centering
    \includegraphics*[width=1\columnwidth]{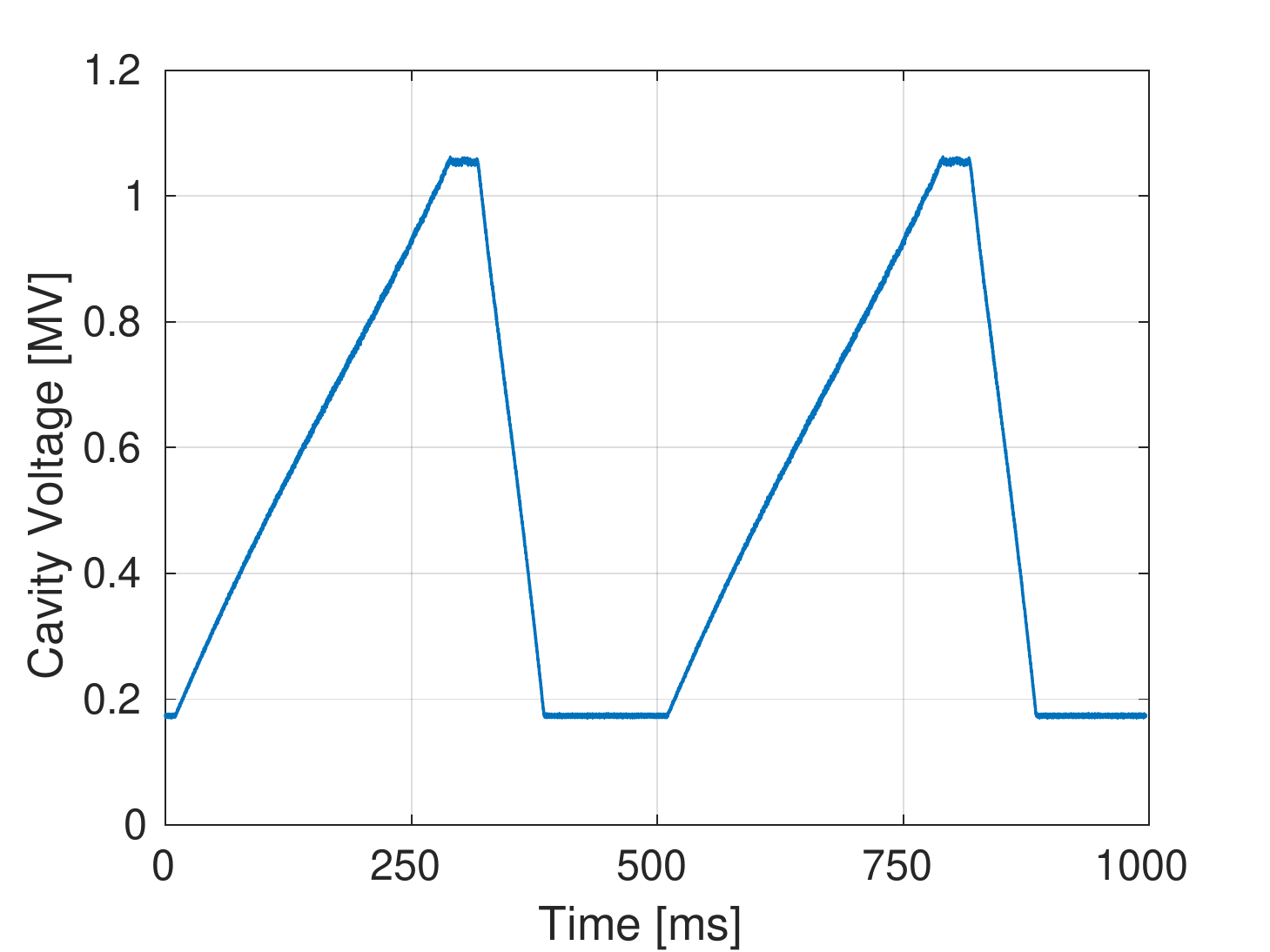}
    \caption{Booster \SI{2}{Hz} cavity voltage ramp.}
    \label{fig:Booster_ramp}
\end{figure}

Th Booster ramp tests at \SI{2}{Hz} cycle are underway and the system is working as expected. Fig.~\ref{fig:Booster_ramp} shows a measurement of the cavity gap voltage during the Booster \SI{2}{Hz} ramp. 

The Slow Loop PI parameters were defined according to a criteria of a overshooting less than \SI{0.5}{dB} on the feedback signal : $kp=10$ and $ki=4100$. Further measurements and tests have to be performed to estimate the transfer function of Booster RF plant to find the best parameters. 

\begin{figure}[!tbh]
    \centering
    \includegraphics*[width=1\columnwidth]{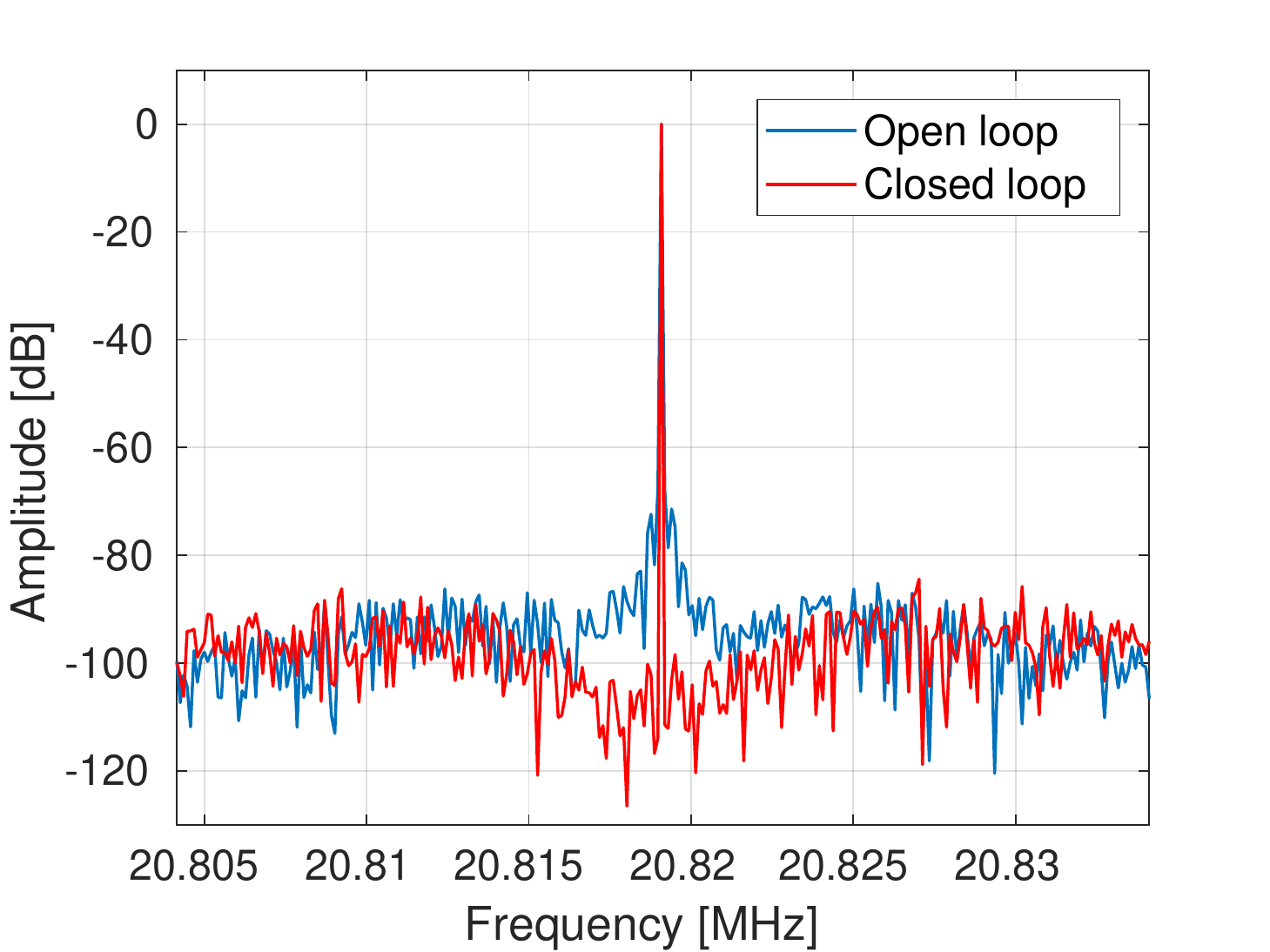}
    \caption{Normalized FFT of raw ADC data in open and closed loop.}
    \label{fig:loop_results}
\end{figure}

Fig.~\ref{fig:loop_results} shows a normalized FFT of the DLLRF ADCs raw data collected by the Fast Data Logger in open and closed loop (slow IQ-loop). These results shows a better frequency response around IF band in closed loop.

\section{CONCLUSION}

The first version of the DLLRF System was assembled in 2017 and the second version is being commissioned in 2019 as the RF subsystems are installed. One of the biggest advantages of using a flexible and modular FPGA based platform is that it allows the addition of extra features to the DLLRF as the operation with the RF systems progresses. 
Sirius Booster RF system is already installed and it is in the final commissioning stage. During the commissioning procedure, the DLLRF has reached the expected specification. Additional tests will be performed to evaluate the maximum system capacity.

\section{ACKNOWLEDGEMENTS}

We would like to acknowledge Angela Salom from CELLS -- ALBA for the valuable contributions and her extensive support on the DLLRF system and Paul Hamadyk from Diamond for sharing their EPICS IOC codes customized for the ALBA's DLLRF. We offer our thanks to our colleagues within Sirius RF and Controls groups for their commitment to get the system ready to work.


%
%
\ifboolexpr{bool{jacowbiblatex}}%
	{\printbibliography}%
	{%
	
	
    } 
%
%


\end{document}